\documentclass[12pt]{article}
\usepackage{psfig}
\usepackage{graphicx}

\begin{document}

{Astrophysics, Vol. 52, No. 2, 2009} \\ \\ \\ \\

\begin{center}

\bigskip

{\large \bf SEARCH FOR COMPANIONS OF NEARBY ISOLATED GALAXIES} \\

{\small O. V. Melnyk$^{1}$, V. E. Karachentseva$^{1}$, I. D. Karachentsev$^{2}$, D. I. Makarov$^{2}$, and I. V. Chilingarian$^{3,4}$}  \\

{\em (1) Astronomical Observatory, Taras Shevchenko Kyiv National University, Ukraine; \\ e-mail: melnykol@gmail.com \\
(2) Special Astrophysical Observatory, Russian Academy of Sciences, Russia\\
(3) Observatoire de Paris-Meudon, France\\
(4) Shternberg Institute of Astronomy, Moscow State University, Russia \\}
\end{center}

{\large \bf Abstract}\\

The radial velocities are measured for 45 galaxies located in the neighborhoods of 29 likely isolated
galaxies in a new catalog. We find that about 85$\%$ of these galaxies actually are well isolated objects.
4$\%$ of nearby galaxies with $V_{LG}<3500$ km/s are this kind of cosmic "orphan". 

\bigskip

\section{Introduction}
Isolated galaxies (IG) are objects which have not been subject to substantial interactions with their nearby
surroundings (another galaxy or groups of galaxies) during their lifetimes. Thus, their observed physical characteristics
are mainly determined by the initial conditions at their formation and by internal evolutionary processes. A
representative sample of isolated galaxies is needed to verify the theory of the origin and evolution of galaxies, and
also as a reference for studies of the properties of galaxies in pairs, groups, and clusters-- that is, in order to understand
the influence of the surroundings on such properties of galaxies as their morphology, chemical composition, and rate
of star formation.

\subsection{Isolated galaxies brighter than 15$^{m}$.7.}

One successful attempt to create this kind of sample is
Karachentseva's Catalog of Isolated Galaxies (CIG) [1]. In compiling it, the following empirically selected criteria
for isolation were used:

\begin{equation}
\label{trivial}
r_{1i}>20a_{i},     
\end{equation}

\begin{equation}
\label{trivial}
4 a_{1 }> a_{i }> a_{1 }/4    
\end{equation}

where the subscripts $1$ and $i$ refer to a fixed galaxy and to its neighbors, respectively. In other words, a galaxy with
angular diameter $a_{1}$  is considered to be isolated if all "significant" neighbors with angular diameters $a_{i }$ in the range
(2) lie a distance $r_{1i}$  from it of at least $20a_{i}$.

A visual survey of the neighborhoods of all the galaxies from the Zwicky CGCG catalog [2] in the O and
E charts of the POSS-1 sky survey showed that the criterion of isolation was satisfied by 1051 galaxies ($m< 15^{m}.7$,
$\delta  > -3\raisebox{1ex}{\scriptsize o}$ , $|b| > 20{\raisebox{1ex}{\scriptsize o}}$ ), or about 4$\%$ of the total number of galaxies in the CGCG.

The fundamental difference between the CIG and the other lists of "individual" galaxies is that neighboring
galaxies are taken into account within a substantial interval of angular diameters (or stellar magnitudes). Therein
lies its major advantage, since ignoring even very slightly fainter neighboring galaxies and a fixed, independent of
the visible magnitude, limit on the angular distance of neighboring galaxies leads to considerable contamination of
the samples of so-called "field galaxies" by members of groups. (See, for example, the work of Stocke [3] and Huchra
and Thuan [4], who examined "field" galaxies from the list of Turner and Gott [5].)

An analysis of the criterion for isolation (Karachentseva [6], Adams et al. [7]) shows that the conditions (1)
and (2) are sufficiently stringent. For galaxies with typical sizes on the order of 20 kpc, with a peculiar velocity
for the field galaxies on the order of 100 km/s, the time to cover a distance equal to 20 diameters is roughly 4 billion
years. This means that the CIG galaxies are isolated for a substantial part of the lifetimes and have not been subjected
to the influence of the surroundings as they evolved. It should be noted that some isolated galaxies could have
experienced a dynamic merger with neighbors in the past. Sometimes cases of this sort can be recognized from a
peculiar shape of a galaxy or an excess of infrared emission.

Over the 35 years since the publication of the CIG, the observational spectral and photometric data base for
isolated galaxies has increased substantially. It suffices to say that the fraction of galaxies in the CIG with measured
radial velocities has increased from 5$\%$ to roughly 95$\%$.

Since the beginning of the century a group of astronomers from Spain, the USA, France, and Italy have been
engaged in the extensive AMIGA project [8]. This project is aimed at studying the physical properties of a reference
sample of the most isolated galaxies selected from the CIG after the initial characteristics have been refined and
supplements on the basis of modern observational data. The results of the papers [9,10] by Verley et al. are of special
interest for this problem.

Verley et al. [9] have conducted an automatic search on the DSS digital sky survey for neighbors with stellar
magnitudes $B<17^{m}.5$ within a circle of radius 0.5 Mpc for each of 950 galaxies in the CIG. After some additional
checks, a catalog of approximately 54000 neighbors was compiled, for 30$\%$ of which the radial velocities are
available in data bases. Without dwelling on the details, here we give the overall statistical result: the neighbors
which were found constitute a population in which objects of the far background predominate. Verley et al. [9,10]
found that the conditions (1) and (2) of the CIG, which eliminate the "significant" neighbors, correspond to an average
difference in radial velocities of the isolated galaxies and the other neighbors of $\sim$ 18000 km/s. The isolation of the
CIG objects was confirmed for more than 80$\%$ of the galaxies that were checked.

Note that in compiling a reference sample of isolated galaxies, the authors of the AMIGA project excluded
100 nearby galaxies in the CIG with $V < 1500$ km/s, since a search for their significant neighbors requires a survey
of large segments of the sky.

\subsection{Isolated galaxies in the Local Supercluster}

The CIG sample, which is limited by the visible magnitude of galaxies, rather than by distance, mainly contains objects with high luminosities from distant volumes. Besides
this, it is important to have a sample of isolated galaxies within the confines of some fixed volume, where dwarf
systems are more completely represented. A sample of this sort has been compiled by Makarov et al. [11] within
the confines of the Local Supercluster and its nearest surroundings. At present, 10403 galaxies are known with radial
velocities relative to the centroid of the Local Group $V_{LG} < 3500$ km/s. In order to exclude groups, a percolation
algorithm was applied to this mass of objects which made it possible to discover about 1300 systems of galaxies with
multiplicities from $n =$ 2 to $n \approx$ 400 [12,13]. In all, about 55$\%$ of the galaxies in this volume were included in these
groups. The only more or less arbitrary parameter for clusterization was the dimensionless quantity $k=M_{T}/M_{25}$, which
indicates the fraction of the total mass of a galaxy that is contained inside its standard radius $R_{25}$. A ratio $k=6$ was
taken for all the galaxies. Evidently, for larger $k$ the relative number of unclustered galaxies is lower. On increasing
$k$ by a factor of 15, we obtained 493 unclustered "field" galaxies, or about 4$\%$ of the entire sample, i.e., the same
relative amount as in the CIG.

In order to discover significant neighbors according to the CIG criterion (1) and (2), we inspected wide
neighborhoods around each of the 493 galaxies in the digital sky surveys DSS and SDSS. We eliminated 17 of the
galaxies from consideration for various reasons: confusion in the velocities or coordinates, close contact with other
galaxies, etc.

This survey showed that 185 of the galaxies have no significant neighbors within a radius of 20$a_{i}$; that is,
they are isolated according to the CIG criterion. Neighbors that were significant in projection were found for the
rest of the galaxies (291); we sought the radial velocities of these in the NED and LEDA data bases. Of these, the
significant neighbors for 34 of the galaxies have close radial velocities. (For 80$\%$ of the cases, the absolute value
of the difference in velocities is less than 100 km/s.) For 111 galaxies the median of the difference in velocities is
$~$10000 km/s, which indicates that the neighbors belong to the distant background.

For the remaining 146 galaxies which were candidate isolated galaxies, from 1 to 13 significant neighbors
without measured radial velocities were found. 88 of the presumed isolated galaxies lie in the northern sky and the
overall number of their significant neighbors is 164. These potential "companions" of northern isolated galaxies also
formed the basis of our observational program for measuring the radial velocities.

\section{Observations and data analysis}

In the second half of 2007 an observational program entitled "A search for the most isolated galaxies in the
Local Supecluster" was carried out at the BTA (6-m telescope at the Special Astrophysical Observatory of the
Russian Academy of Sciences). This yielded images and spectra of 45 significant neighbor galaxies for 29 candidate
nearby isolated galaxies. The observations were made over 6 partial nights: August 6, October 7, and October 9-
12, 2007. The spectra were obtained using the universal luminosity reducer SCORPIO [14] with a long slit and a
VPHG550G grating, at the primary focus.

The spectral resolution was 10 $\AA$/pixel (CCD EEV42-40) over the wavelength range 3100-7300 $\AA$ and the
central wavelength was 5100 $\AA$. The visible magnitudes of the observed objects lay in the range $m$ = 17-18 and
the exposure times were from 300 to 900 seconds.

The spectra obtained from the CCD were reduced using the program package LONG in the standard MIDAS
system. The radial velocity was determined in two ways: from the absorption lines by fitting the observed spectrum
with synthetic spectra from the PEGASE.HR program package [15] and from the emission lines, as an average value
calculated from individual lines.

The columns of Table 1 list the following data:
(1) the name of the main galaxy, (2) its heliocentric radial velocity in km/s according to the NED, (3) the name
or coordinates of the significant neighboring galaxy, (4) the mutual angular separation between the main galaxy and
the significant neighbor in minutes of arc, and (5) our measured radial velocity for the significant neighbor in
km/s. The typical error in the velocity measurements is 25 km/s.

\section{Discussion}

Figure 1 is a plot of the distribution of the difference in the radial velocities of the 45 significant (in
projection) neighbors relative to the 29 proposed isolated galaxies. The velocity difference $V_{i} -V_{1}$ has a wide range
extending to 40000 km/s with a median of 9400 km/s. Only 4 of the galaxies out of the 29, UGC3234, UGC10806,
UGC11251, and UGC12504 have neighbors whose radial velocities lie with $\pm$ 200 km/s of the galaxy considered to
be isolated. The mutual linear distances for these pairs are small, and equal to 47 kpc (UGC3234 and 0504+16), 11
kpc (UGC 10806 and PGC 167108), 159 kpc (UGC11251 and 1827+38), and 79 kpc (UGC 12504 and PGC 2773857),
with an average value of 74 kpc.

The closest example of these four cases, the galaxy UGC10806 with a radial velocity of +932 km/s, is shown
in Fig. 2 which reproduces the $8'\times15'$ neighborhood of the pseudoisolated galaxy from the Palomar survey of the sky
(POSS-II, R). The two faint "companions" to the west of UGC10806 have radial velocities of +7354 and
+7569 km/s; that is, they are objects in the distant background. The physical companion to the east of UGC10806
has a velocity of +936 km/s. Its luminosity is $3^{m}$.3 fainter than the galaxy being examined for isolation.
On the average, in terms of luminosity the 4 physical companions are fainter than the candidate isolated
galaxies by a factor of 7; that is, tidal perturbations from them are not very great. With an average velocity difference
of 66 km/s and an average luminosity of $1.9\times 10^{9} L_{\odot}$ , the ratio of the orbital mass to the luminosity for these pairs
is 130$M_{\odot} /L_{\odot}$  , which is consistent with the physical nature of these pairs.

Our observations show that the fraction of truly isolated galaxies was $25/29 = 86\%$. Despite the small
statistics, this estimate is in good agreement with data comparing the radial velocities of galaxies from the NED and
LEDA data bases, where spatial isolation was not confirmed for only 12$\%$ of the "field" galaxies with $V_{LG} <$ 2500
km/s. Recall that the data of Refs. 9 and 10 confirm the spatial isolation of the CIG objects [1] for more than 80$\%$
of the galaxies.

The sample of 493 candidate nearby isolated galaxies, therefore, does contain a quite high fraction of truly
isolated objects and can be used for analyzing the physical properties of galaxies in regions with an extremely low
density of matter. \\ \\

Table 1. Radial Velocities of Neighbors of Isolated Galaxies

\begin{tabular}{|l|l|l|l|l|}
\hline
Name of & $V_{h}$, km/s & Name or coordinates & $r_{1i}$ & $V_{h},$ \\
main galaxy &   & of neighboring galaxy & (ang. min) & km/s \\
\hline
1& 2 & 3 & 4 & 5 \\
\hline
UGC12921 &2449 & 2MASXJ00013040+7726001 & 10.9 &19518 \\
NGC0048 &1776 & 2MASXJ00142204+4816525 & 4.3 & 4962\\
UGC00199 &1800 & 002025.5+125445 & 6.8 & 42692\\
UGC00313 & 2085 & 003124.9+061527 & 3.1 & 31205\\
 &  & 003215.5+062037 & 14.3 & 10207\\
 &  & 003222.3+055400 & 21.8 & 11420\\
UGC00392 & 1403 & 2MASXJ00354608+8313586 & 8.5 & 7860\\
 & & 2MASXJ00510454+8312244 & 18.6 & 11926\\
UGC00578 & 1471 & 005624+395122 & 1.6 & 28802\\
& & 2MASX J00564535+3949592 & 4.7 & 17066\\
& & 005647.5+395520 & 7.2 & 10884\\
& & 2MASXJ00565543+3953101 & 7.5 & 21866\\
UGC00614 & 2354 & PGC2066499 & 25.2 & 18381\\
UGC01054 & 2663 & 012826+342153 & 4.7 & 5063\\
& & 012827+335750 & 21.8 & 4507\\
UGC01211 & 2408 & 014353.3+135229 & 3.6 & 13581\\
UGC01207 & 1256 & 015112+820650 & 8.1 & 4689\\
UGC03045 & 1387 & PGC1627761 & 5.9 & 3771\\
UGC03199 & 2100 & 045443+015202 & 13.4 & 9291\\
UGC03234 & 1402 & 050403+162435 & 9.1 & 1267\\
UGC03394 & 1821 & 2MASXJ06012922+5608590 & 27.9 & 13184\\
& & 2MASXJ06081506+5600244 & 30.2 &11289\\
UGC03476 & 469 & 062957+331900 & 6.3 & 7269\\
UGC03475 & 487 & MCG+07-14-001 & 33.8 & 5320\\
UGC03485 & 1285 & 2MASXJ06354709+6543083 & 8.7 & 26532\\
UGC03501 & 449 & 2MASXJ06390063+4930357 & 15.5 & 5884\\
PGC2807122 & 451 & 064814+472340 & 7.5 & 16412\\
UGC3600 & 412 & 065557+391650 & 11.2 & 25978\\
NGC6339 & 2108 & MCG+07-35-062 & 3.0 & 8711\\
& & 171803.5+405540 & 11.5 & 10804\\
& & 182710+380250 & 8.3 & 9421\\
UGC10806 & 932 & 171812.0+495615 & 7.7 & 7569\\
& & PGC167103 & 3.3 & 7354\\
& & PGC167108 & 3.1 & 936\\
UGC10892 & 1927 & 2MASXJ17293881+7410216 & 5.5 & 20232\\
& & 173059.5+740910 & 9.6 & 20129\\
NGC6434 & 2483 & 2MASXJ17353612+7203453 & 5.8 & 16504\\
& & 2MASXJ17364223+7155241 & 10.0 & 29694 \\
UGC11251 & 2334 & 182710+380250 & 18.0 & 2225\\
& & 2MASXJ18283937+3803123 & 1.7 & 23864\\
& & 182840+380150 & 1.0 & 23987\\
NGC6675 & 2495 & PGC2159396 & 3.4 & 19436\\
NGC6757 & 2380 & MCG+09-31-018 & 29.4 & 16094\\
UGC11496 & 2121 & UGC11476 & 44.4 & 5385\\
UGC12504 & 2436 & PGC2773857 & 8.4 & 2421\\
\hline
\end{tabular}

\begin{figure}[t]
\includegraphics[angle=0, width=0.8 \textwidth]{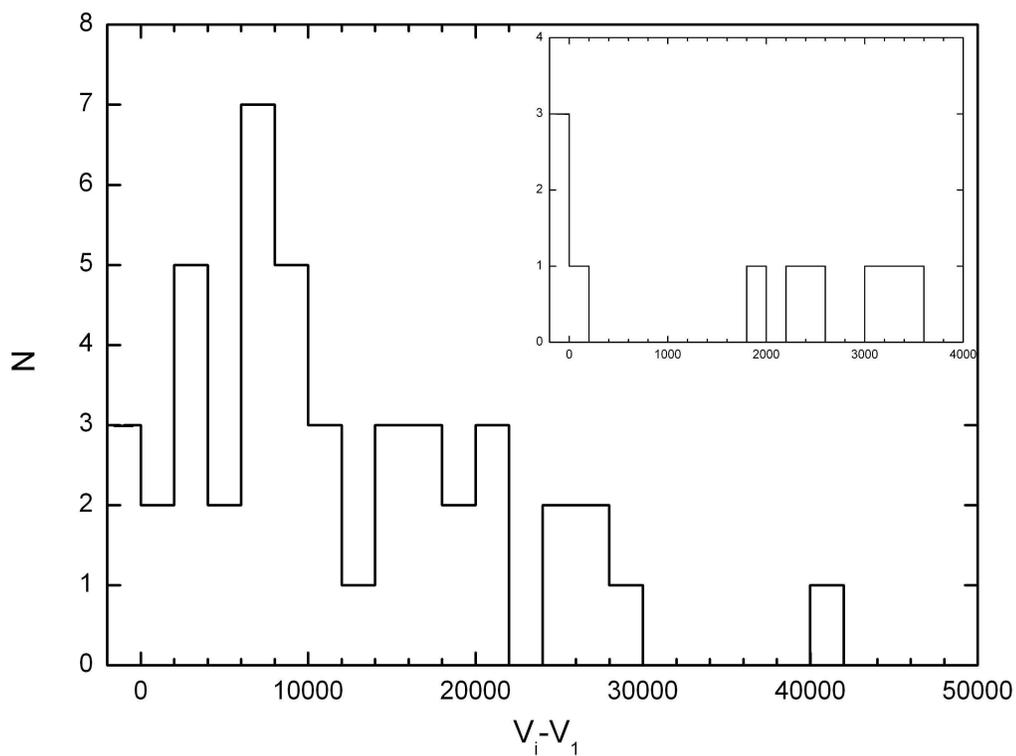}
\caption{The distribution of the velocities $V_{i}$ of 45
"significant" neighbors with respect to the velocities $V_{1}$  of
the presumed isolated galaxies (km/s)}
\end{figure}

\begin{figure}[t]
\includegraphics[angle=0, width=0.8 \textwidth]{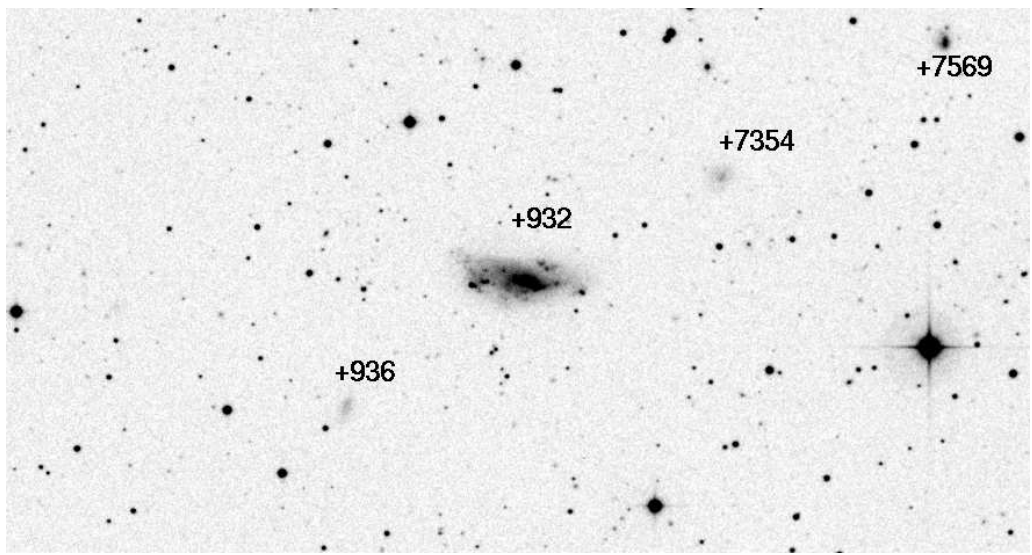}
\caption{The nominally isolated galaxy UGC10806 (center) and its
neighborhood ($15'\times 8'$) reproduced from the DSS (POSS-II, R). North is
upward, east to the left. The radial velocities of UGC10806 and three of its
neighbors are indicated in the figure.}
\end{figure}

\clearpage

\bigskip

{\bf ACKNOWLEDGMENTS.} This work was partially supported by the "Cosmomicrophysics" complex program of the Ukrainian National
Academy of Sciences and by grants from the Russian Foundation for Basic Research RFFI 07-02-00005 and RFFI-DFG 06-02-04017.

{}
\end{document}